\documentclass[11pt,a4paper]{article}
\usepackage{jcappub}
\usepackage{graphicx}
\usepackage[lofdepth,lotdepth]{subfig}
\usepackage{dcolumn}
\usepackage{bm}

\newcommand {\be}{\begin{equation}}
\newcommand {\ee}{\end{equation}}
\newcommand {\ba}{\begin{eqnarray}}
\newcommand {\ea}{\end{eqnarray}}

\title{Implications of  the Pseudo-Dirac Scenario for Ultra High Energy Neutrinos from GRBs}

\author[a]{Arman Esmaili}
\author[b]{, Yasaman Farzan}
\emailAdd{aesmaili@ifi.unicamp.br}
\emailAdd{yasaman@theory.ipm.ac.ir}
\affiliation[a]{Instituto de Fisica Gleb Wataghin - UNICAMP, 13083-859, Campinas, SP, Brazil}
\affiliation[b]{School of physics, Institute for Research in Fundamental Sciences (IPM), P.O.Box 19395-5531, Tehran, IRAN}

\abstract{The source of Ultra High Energy Cosmic Rays (UHECR) is still an unresolved mystery. Up until recently, sources of Gamma Ray Bursts (GRBs) had been considered as a suitable source for UHECR. Within the fireball model, the UHECR produced at GRBs should be accompanied with a
neutrino flux detectable at the neutrino telescope such as IceCube. Recently, IceCube has set an upper bound on the neutrino flux accompanied by GRBs about 3.7 times below the prediction. We
investigate whether this deficit can be explained by the oscillation of the active neutrinos to sterile neutrinos {\it en route} from the source to the detectors within the pseudo-Dirac scenario. We then discuss the implication of this scenario for diffuse supernova relic neutrinos.}

\begin{document}

\maketitle

\section{Introduction\label{intro}}

One of the challenges in cosmic ray physics is identifying the source of Ultra High Energy Cosmic Rays (UHECRs) with energy $>10^{18}$~eV. Simple calculation of the energy budget and rate of  astrophysical explosions shows that the transient Gamma Ray Bursts (GRBs)  can accommodate the observed flux of UHECRs~\cite{Waxman:1995vg}. The GRBs are energetic explosions with luminosity $\sim10^{51}$~erg/s at cosmological distances which have been observed by various satellites via their gamma ray emissions at a  rate $\sim$~2/day.

Within the fireball model of the GRB explosion ({\it see}~\cite{Waxman:2003vh} for a review on this model), the decay of the photo-pions produced by accelerated protons in the expanding shock wave generates a flux of accompanying neutrinos. Based on the observed flux of UHECRs, the expected flux of neutrinos is $E_\nu^2dN_\nu/dE_\nu\sim 5\times 10^{-9}\,{\rm GeV}{\rm cm}^{-2}{\rm s}^{-1}{\rm sr}^{-1}$ in the energy range $\sim 100\,{\rm TeV}-10\,{\rm PeV}$ (the so-called Waxman-Bahcall flux~\cite{Waxman:1997ti}). On the other hand within the fireball model, the neutrino flux emitted from each GRB can be calculated from the observation of gamma ray energy spectrum. Normalizations based on the cosmic ray flux and the gamma ray flux both result in approximately the same prediction for the neutrino flux. The Waxman-Bahcall flux leads to $\sim$ 10 events per year
in a km$^3$-scale neutrino telescope.

Recently, the IceCube experiment published the result of the analysis of data taken during the construction of detector, that is IC40 and IC59 \cite{Abbasi}. The combined analysis of the data
does not show any neutrino signal correlated with the observed GRBs during the data-taking period. The limit of IceCube on the neutrino flux is $\sim 3.7$ times smaller than the prediction \cite{Abbasi}. From this limit, the IceCube collaboration has concluded that the GRBs cannot account for the
bulk of the UHECRs in the universe \cite{Abbasi}. However, this conclusion can be questioned in three ways: (1) As shown in \cite{Winter}, within the fireball model, uncertainties are so large that the prediction for the neutrino flux can be lower than the standard Waxman-Bahcall prediction by up to a factor of ten (see also~\cite{Murase:2005hy}). (2) In \cite{Dar}, it is discussed that within the cannonball model, the sources of GRBs can account for the UHECR without violating the neutrino flux bound because the cannonball model does not predict a sizable neutrino flux. (3) Finally, it is possible that some beyond standard model effects reduce the flux after production (see {\it e.g.} \cite{Barranco}). In this letter, we entertain the third possibility.

Although overwhelming evidence has been gathered showing that there are at least two massive neutrino mass eigenstates, we do not still know whether neutrinos are of Majorana type or of Dirac
type. An interesting possibility is to have pseudo-Dirac neutrino mass scheme~\cite{Wolfenstein:1981kw}, where the dominant contribution to the neutrino mass comes from the Dirac mass term, $m_D$, with a small correction from the Majorana mass term, $m_M$ such that $m_M\ll m_D$. Notice that the presence of the Dirac term requires right-handed sterile neutrinos. Within the
pseudo-Dirac scenario, there is a small mass splitting between active and sterile neutrinos making the oscillation between active and sterile components in principle possible. For very small $m_M$, the oscillation length will be too large for the atmospheric and solar neutrinos to oscillate to sterile ones. However for neutrinos traveling over cosmological distances, the baseline will be large enough for active-sterile oscillation to take place~\cite{pD}. In this paper, we investigate this
possibility to relax the conflict between IceCube bound on the neutrinos from GRBs and the expected neutrino flux from GRBs as the origin of UHECR.

The paper is organized as follows: In section~\ref{sec:scenario}, we briefly discuss the pseudo-Dirac scenario for neutrino masses. In section~\ref{sec:grb}, we investigate the possible depletion of GRB neutrino flux due to the presence of tiny active-sterile mass-squared difference predicted in pseudo-Dirac scenario. Section~\ref{sec:sn} is devoted to the implications of pseudo-Dirac scenario for supernova neutrinos, discussing both single source and diffuse fluxes. The conclusions are summarized in section~\ref{sec:conc}.

\section{\label{sec:scenario}Pseudo-Dirac Scenario}

In general, a neutral fermion such as neutrino can have two types of mass term: (1) Dirac mass term, $m_D \bar{\nu}\nu$; (2) Majorana mass term, $\frac{1}{2}m_M(\nu^T\mathcal{C} \nu-\nu^\dagger \mathcal{C} \nu^*)$, where $\mathcal{C}$ is the charge conjugation operator. Dirac mass term requires the existence of a right-handed neutrino: $m_D\bar{\nu}\nu=m_D(\bar{\nu}_R\nu_L+\bar{\nu}_L\nu_R)$. If there is only Dirac mass term, the neutrino mass is said to be purely of Dirac type and the lepton number is conserved. For the case of pure Dirac neutrino mass term, the mass eigenstates will be the superposition of Majorana states with maximal mixing and the mass eigenvalues will be equal. To illustrate this point, consider one generation pure Dirac mass term. In this case, the two Majorana states $\nu^+\equiv (\nu_L+~\nu_R^c)/\sqrt{2}$ and $\nu^-\equiv(\nu_L-\nu_R^c)/\sqrt{2}$ will be mass eigenstates with masses equal to $m^+=m_D$ and $m^-=-m_D$, respectively. The two states $\nu_R$ and $\nu_L$ are degenerate in mass, which taking into account that their masses are lighter than $m_Z/2$, implies that  $\nu_R$ has to be sterile. Notice that despite  the maximal active-sterile mixing, there will be no oscillation between active and sterile neutrinos because the squares of the masses are exactly equal.

Let us now suppose, through a lepton number violating process, the right-handed neutrinos obtain a very small Majorana mass, such that $m_M\ll m_D$ (which justifies the nomenclature ``pseudo-Dirac"). The smallness of $m_M$ can be justified within the `t Hooft criterion as in the limit $m_M\to 0$, the lepton number is conserved. In this case the degeneracy between $(m^+)^2$ and $(m^-)^2$ will be lifted and the active-sterile oscillation can take place. Generalization to three generations of neutrinos is straightforward. In this case the generic mass term consisting of both Dirac and Majorana mass terms, can be written as $\mathcal{L}_{\rm mass}=-\frac{1}{2} \overline{\Psi^c} M \Psi$, where $\Psi=(\nu_{L 1}, \nu_{L 2}, \nu_{L 3}, \nu_{R 1}^c, \nu_{R 2}^c,\nu_{R 3}^c)^T$, and the $6\times6$ mass matrix is given by
\ba
M=\left(\begin{matrix} 0 & m_D^T\cr m_D &m_M^\ast\end{matrix} \right)~,
\ea
where $m_D$ and $m_M$ are the $3\times3$ Dirac and right-handed Majorana mass matrices, respectively. It is convenient to go to the mass basis where $m_D$ is diagonal:
$$m_D={\rm diag}\, (m_1,  m_2 ,m_3)~,$$
where $m_2=\sqrt{m_1^2+\Delta m_{\rm sol}^2}$ and $m_3=\sqrt{m_1^2+\Delta m_{\rm atm}^2}$, with $\Delta m_{\rm sol}^2$ and $\Delta m_{\rm atm}^2$, respectively, being the solar and atmospheric mass-squared differences obtained in neutrino oscillation phenomenology. In this basis, $m_M$ is a general symmetric matrix with $m_M\ll m_D$ (order relation applies to the nonzero  elements of the matrices). The mass eigenstates are given by
\begin{eqnarray}
\nu^+_{i} &=& \frac{\nu_{Li}+\nu_{Ri}^c}{\sqrt{2}}+\sum_{j\ne i}
(\alpha_{ij}^+ \nu_{L j} +\beta_{ij}^+ \nu_{R j}^c)~, \cr \nu^-_{i} &=& \frac{\nu_{Li}-\nu_{Ri}^c}{\sqrt{2}}+\sum_{j\ne i}
(\alpha_{ij}^- \nu_{L j} +\beta_{ij}^- \nu_{R j}^c)~,
\end{eqnarray}
where the coefficients $\alpha_{ij}^\pm,\beta_{ij}^\pm \sim m_M/m_D\ll 1$ and the following orthogonality relations apply to the mass eigenstates: $\langle \nu^+_{i}|\nu^-_{i} \rangle=0$ and $\langle \nu^-_{i}|\nu^-_{j}\rangle=\langle\nu^+_{i}|\nu^+_{j}\rangle=0$ (for $i\ne j$). The mass eigenvalues corresponding to $\nu^+_{ i}$ and $\nu^-_{i}$ are, respectively, $(m^+_{i})^2=m_i^2+\Delta m_i^2/2$ and $(m^-_{i})^2=m_i^2-\Delta m_i^2/2$, where $\Delta m_i^2 \sim m_Dm_M\ll \Delta m_{\rm sol,atm}^2$~.

\begin{figure}[t]
\begin{center}
\includegraphics[width=0.75\textwidth]{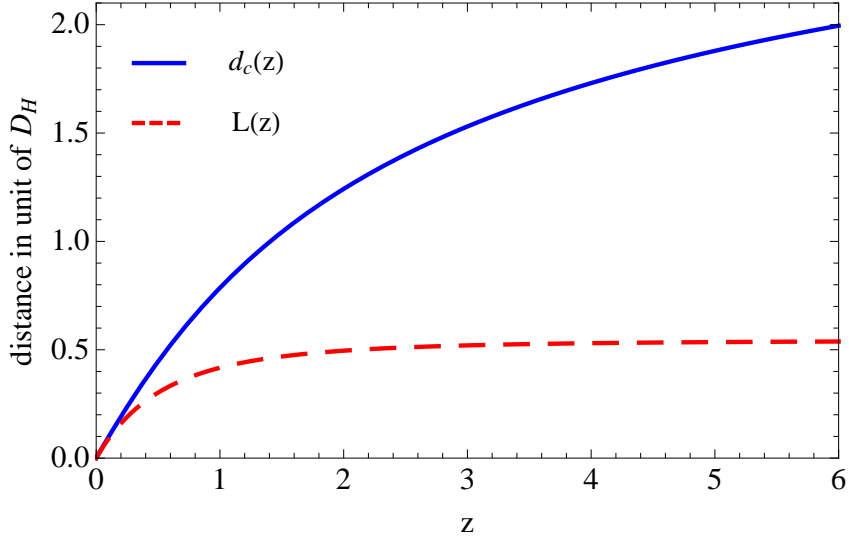}
\caption{\label{fig:dis}The dashed red curve shows the distance $L(z)$, defined in Eq.~(\ref{distance}), as a function of redshift. The solid blue curve corresponds to the comoving distance defined in Eq.~(\ref{eq:comoving}). The vertical axis is in the unit of $D_H=c/H_0=4.15$~Gpc.}
\end{center}
\end{figure}

In the pseudo-Dirac scenario the probability of the $\nu_\alpha \to \nu_\beta$ oscillation among the active neutrinos ($\alpha,\beta=e,\mu,\tau$) can be written as:
\begin{equation} \label{exact}
P_{\alpha\beta}=\frac{1}{4}\left| \sum_{j=1}^3 U_{\alpha j}\left\{
e^{i\Phi_j^+}+e^{i\Phi_j^-} \right\} U_{\beta j}^\ast \right|^2~,
\end{equation}
where $U_{\alpha i}$ are the elements of conventional $3\times3$ neutrino mixing matrix (PMNS matrix). The phases $\Phi_j^\pm$ for a neutrino with momentum $k$ at the Earth is given by
\be \label{phase}
\Phi_j^{\pm}=\int_t^{t_0} \left[ (m^{\pm}_{j})^2+k^2
\left(\frac{a(t_0)}{a(t^\prime)}\right)^2\right]^{1/2}{\rm d}t^\prime~,
\ee
where $a$ is the time-dependent scale factor  appearing in the metric ({\it i.e.,}
$g_{\mu \nu} = ( -1 , a^2 , a^2 , a^2)$). The time of neutrino emission and  the present time are respectively denoted by $t$ and $t_0$. For baselines $\ll 100$~Mpc, $a(t)/a(t_0)\to 1$ and  we therefore obtain the standard result: $\Phi_j^{\pm}=\sqrt{ k^2+(m^{\pm}_{j})^2} (t-t_0)$. For the baselines $4\pi E_\nu/\Delta m_{\rm sol}^2\ll L \ll 100$~Mpc, the oscillatory terms originating from interference of different $j$ ({\it i.e.}, oscillation between pairs of states induced by $\Delta m_{\rm sol}^2$ and $\Delta m_{\rm atm}^2$ scales) will average out and from Eqs.~(\ref{exact}) and (\ref{phase}) we obtain
\begin{equation}
\label{short}
P_{\alpha\beta}=\sum_{j=1}^3 |U_{\alpha j}|^2 |U_{\beta j}|^2
\cos^2 \left( \frac{\Delta m_j^2 L}{4E_\nu} \right)~,
\end{equation}
where $\Delta m_j^2 \equiv (m_j^+)^2-(m_j^-)^2$ is the tiny mass-squared difference within the $j$-th pair of active and sterile states. From now on, we assume equal $\Delta m_j^2$ for all the pairs; {\it i.e.}, $\Delta m_j^2\equiv\Delta m^2$ for $j=1,2,3$;  we however keep the symbol $\Delta m_j^2$ with its subscript ``$j$''. We will discuss the consequences of relaxing this assumption in the last section. By using Eq.~(\ref{short}) in the analyses of oscillation experiments, it is possible to extract
or derive limits on the parameter $\Delta m_j^2$. Notice that the oscillation length corresponding to $\Delta m_j^2$ is given by
\begin{equation}\label{eq:oscL}
L_{\rm osc}\equiv \frac{4\pi E_\nu}{\Delta m_j^2}\simeq 80~{\rm kpc}~\left( \frac{10^{-12}~{\rm eV}^2}{\Delta m_j^2} \right) \left( \frac{E_\nu}{{\rm PeV}} \right)~,
\end{equation}
which means that for the solar neutrinos with $E_\nu\sim~{\rm MeV}$, oscillation length will be of the order of 1 a.u.~for $\Delta m_j^2\sim 10^{-12}~{\rm eV}^2$. The detailed analysis of the solar neutrinos gives the strongest bound on $\Delta m_j^2$, which is $\Delta m_j^2<1.8\times 10^{-12}~{\rm eV}^2$ at $3\sigma$ level~\cite{solar}. For neutrinos coming from cosmological distances $\sim$~Gpc with energy $E_\nu\sim10$~PeV, the baseline is much larger than the oscillation length ($L\gg L_{\rm osc}$) for $\Delta m_j^2>10^{-16}~{\rm eV}^2$. However, for such distances the exact formulae in Eqs.~(\ref{exact}) and (\ref{phase}) have to be used instead of Eq.~(\ref{short}). After averaging out the oscillatory terms induced by the solar and atmospheric mass-squared differences, the oscillation probability in Eq.~(\ref{exact}) can be written as
\begin{equation}\label{shortexact}
P_{\alpha\beta}=\sum_{j=1}^3 |U_{\alpha j}|^2 |U_{\beta j}|^2 \cos^2 \left( \frac{\Delta \Phi_j}{2} \right) \ \ {\rm where} \ \ \Delta \Phi_j\equiv \Phi_j^+-\Phi_j^-~.
\end{equation}
Notice that as long as we are in the ultrarelativistic limit with $k\gg m^\pm_{j}$, the phase difference $\Delta \Phi_j$ remains proportional to $\Delta m_j^2$. Thus for $\Delta m_j^2 \to 0$, we recover the standard formula for complete loss of coherence: $P_{\alpha \beta}= \sum_i |U_{\alpha i}|^2|U_{\beta i}|^2$. From Eq.~(\ref{phase}), the phases can be calculated as:
\begin{eqnarray}
\Phi_j^\pm &= &\int_{t}^{t_0} \frac{k}{a(t^\prime)} \left[ 1+ \left(\frac{m_j^\pm a(t^\prime)}{k}\right)^2  \right]^{1/2} {\rm d}t^\prime \nonumber \\
& \simeq &\int_{t}^{t_0} \frac{k}{a(t^\prime)}~{\rm d}t^\prime + \frac{(m_j^\pm)^2}{2}\int_{t}^{t_0} \frac{a(t^\prime)}{k}~{\rm d}t^\prime~.
\end{eqnarray}
Thus,
\begin{equation}
\Delta \Phi_j\equiv\Phi_j^+-\Phi_j^- =  \frac{\Delta
m_j^2}{2}\int_{t}^{t_0} \frac{a(t^\prime)}{E_\nu}~{\rm d}t^\prime~,
\end{equation}
where $E_\nu=k$ is the neutrino energy at Earth which is related to the neutrino energy at source by $E_\nu=E_\nu^0/(1+z)$. Let us define $a(t_0)=1$. Using  $a(t)=1/(1+z)$  and  $|dz/dt|=H_0 (1+z) \sqrt{\Omega_m (1+z)^3+\Omega_\Lambda}$, we obtain
\begin{equation}\label{eq:deltaphi}
\Delta\Phi_j =  \frac{\Delta m_j^2}{2E_\nu} D_H
\int_{0}^{z} \frac{{\rm d}z^\prime}{(1+z^\prime)^2\sqrt{\Omega_m
(1+z^\prime)^3+\Omega_\Lambda}}~,
\end{equation}
where $D_H=c/H_0$ is the Hubble length with $H_0=71~{\rm km}~{\rm s}^{-1}~{\rm Mpc}^{-1}$ denoting the present time Hubble expansion rate. $\Omega_m=0.27$ and $\Omega_\Lambda=0.73$ are the matter and dark energy density of universe, respectively. Thus, for the phase difference appearing in Eq.~(\ref{shortexact}), we obtain
\begin{equation}\label{phase1}
\Delta\Phi_j =  \frac{\Delta m_j^2}{2E_\nu} L(z)~,
\end{equation}
where
\begin{equation}\label{distance}
L(z) = D_H\int_{0}^{z} \frac{{\rm d}z^\prime}{(1+z^\prime)^2\sqrt{\Omega_m (1+z^\prime)^3+\Omega_\Lambda}}~.
\end{equation}
Notice that $L(z)$ is different from the familiar quantities such as comoving or luminosity distance and it already includes the redshift evolution of oscillation phase. In order to clarify the difference, in
Figure~\ref{fig:dis} we show the distance $L(z)$ as a function of redshift by the dashed red curve. For comparison, we have also shown the comoving distance $d_c(z)$ by solid blue curve [see
Eq.~(\ref{eq:comoving})]. For a given $E_\nu$, the asymptotic behavior of $L(z)$ at high redshifts, sets a lower limit on $\Delta m_j^2$ that can be probed by this method. This limit is of the order of $10^{-17}~{\rm eV}^2(E_\nu/{\rm PeV})$. From this formula, we observe that for the purpose of probing smaller $\Delta m_j^2$, lower energy neutrinos have to be employed. A suitable choice can be supernova neutrinos with $E_\nu$ of the order of 10~MeV which can be sensitive to $\Delta m_j^2$ down to $10^{-25}~{\rm eV}^2$. Discussion about the neutrino oscillation in curved space-time can be found also in~\cite{Weiler:1994hw}.

Let us now discuss two cases $L\gg L_{\rm osc}$ and $L\sim L_{\rm osc}$ one by one.

\begin{itemize}

\item
In case that the baseline $L$ is much larger than $L_{\rm osc}$, the phases $\Delta \Phi_j$ will be very large. Since the spectrum of neutrinos from cosmological sources is continuous and the
energy resolution of the detector is finite, regardless of the spatial distribution of the sources, the oscillatory terms induced by $\Delta m_j^2$ will average out. Taking the average of the cosine term in Eq.~(\ref{shortexact}), we obtain \be\label{eq:half} P_{\alpha \beta}= \frac{1}{2}\sum_i |U_{\alpha
i}|^2|U_{\beta i}|^2~. \ee Thus, for the case $L\gg L_{\rm osc}$ we expect a suppression factor $1/2$ of the neutrino flux with respect to the standard oscillation result due to the presence of tiny mass-squared differences $\Delta m_j^2$ predicted in the pseudo-Dirac scenario. In particular $\sum_\beta P_{\alpha\beta}=1/2$, which reflects the fact that within the pseudo-Dirac scenario, half of the active neutrinos are converted to sterile neutrinos over large enough distances.

In the limit $L\gg L_{osc}$, this factor of half suppression turns out to be very robust. In particular, as we discuss below, possible matter effects at the source do not change it. If the effective potentials due to the matter effects ({\it i.e.,} $V_C=\sqrt{2} G_F n_e$ and $V_N=-\sqrt{2}G_F n_n/2$) are much larger than $\Delta m_j^2/2 E_\nu$, the effective active-sterile mixing will be suppressed and the conversion from active to sterile neutrinos will be therefore suppressed. In order for the matter effects to play a significant role in the oscillation, the following two conditions have to be fulfilled: (1) $\int V_N~ dt\sim \int V_C~ dt\sim 1 \ {\rm or} >1$; (2) $V_N\sim V_C \sim \Delta m_j^2/2 E_\nu$ (which is $\ll \Delta m_{\rm sol}^2/2 E_\nu$). The first condition, which has been studied in detail in \cite{Cecilia}, is independent of the neutrino mixing and mass splitting. In a region where the second condition is fulfilled, there will not be any chance of conversion between $\nu^\pm_{i}$ and $\nu^\pm_{j}$ (with $i\ne j$) due to matter effects, as the splitting between $m^+_{i}(\simeq m^-_{i})$ and $m^+_{j}(\simeq m^-_{j})$ are  given by $(\Delta m_{sol}^2~{\rm or}~\Delta m_{atm}^2)/(m_i+m_j)$ and are therefore too large. However, due to matter effects $\nu_{L i}\simeq (\nu^+_{i}+\nu^-_{i})/\sqrt{2}$ can be converted to $\cos \theta_i \nu^+_{i}+\sin \theta_i e^{i\alpha_i} \nu^-_{i}$, where $\theta_i$ and $\alpha_i$ depend on the matter profile in the medium through which the neutrino is passing. Suppose after exiting the matter, the neutrino traverses a large distance in vacuum. Then,
$$P(\nu_{Li} \to \nu_{Li})\simeq \left| \frac{\cos \theta_i e^{i\phi_i^+}+\sin \theta_i   e^{i\alpha_i}e^{i\phi_i^-}}{\sqrt{2}}\right|^2~.$$
Thus, in the averaging limit  $P(\nu_{Li} \to \nu_{Li})=(\cos^2\theta_i+\sin^2\theta_i)/2=1/2$. Now, consider a general state $\nu_{int}=\sum_i s_i \nu_{Li}$ with $\sum_i s_i^2=1$. The $\nu_{int}$ state might correspond to  a flavor state or might be a coherent mixture of them after passing denser regions. Suppose $\nu_{int}$ arrives at such a region and transverses it and then travels a long distance $L\gg L_{\rm osc}$ in vacuum. The oscillation probability to $\nu_\mu$ will then be $P(\nu_{int}\to \nu_\mu)=\sum_i |U_{\mu i}|^2 \frac{s_i^2}{2}$. This is the same results (suppression factor of two) that we would obtain without the matter effects on the active-sterile conversion.

\item
For the baselines comparable with the oscillation length, $L\sim L_{\rm osc}$, the $\cos^2 (\Phi_j/2)$ in Eq.~(\ref{shortexact}) does not necessarily average to $1/2$ for an individual source and it can be smaller, leading to a strong suppression of active neutrino flux. However, for a flux of neutrinos originating from various sources with different baselines, one should also average over the different baselines. The suppression will then depend on the redshift distributions of the sources. In particular, if the neutrino sources are clustered around one or a few values of redshift $z$, for a given $E_\nu$, the suppression factor can be made arbitrarily small by choosing the appropriate range of $\Delta m_j^2$. In order to quantify the significance of the suppression of neutrino flux, we define the effective suppression factor, $S_{\rm eff}$, as a function of $\Delta m_j^2$ and neutrino energy at the detector, $E_\nu$, in the following way:

\begin{equation}\label{eq:sup}
S_{\rm eff}(\Delta m_j^2,E_\nu)= \frac{\sum_k \left
\langle\cos^2\left(\frac{\Delta\Phi_j(z_k,E_\nu^0)}{2}\right)
\right\rangle \frac{dN_\nu
(z_k,E_\nu^0)}{dE^0_\nu}\frac{(1+z_k)}{\left[d_c(z_k)\right]^2}}{\sum_k
\frac{dN_\nu (z_k,E_\nu^0)}{dE^0_\nu}\frac{1+z_k}
{\left[d_c(z_k)\right]^2}}~,
\end{equation}

where  $[dN_\nu(z_k,E_\nu^0)/dE^0_\nu]dE^0_\nu$ is the total number (fluence) of neutrinos  from a source at $z_k$ and with an energy between $E_\nu^0=E_\nu(1+z_k)$ and $E_\nu^0+dE_\nu^0$. The sum is over the individual neutrino sources at redshifts $z_k$. The symbol $\langle X\rangle$ shows the average of the quantity $X$ over the energy resolution of the detector. The factors of $(1+z_k)$ in the numerator and denominator come from the redshift of the energy interval: $dE_\nu^0=dE_\nu(1+z_k)$. The factor $1/d_c^2(z_k)$ is the comoving distance which accounts for the geometrical factor which enters in the calculation of neutrino flux:
\begin{equation}\label{eq:comoving}
d_c(z)=D_H \int_{0}^{z} \frac{{\rm d}z^\prime}{\sqrt{\Omega_m(1+z^\prime)^3+\Omega_\Lambda}}~.
\end{equation}

For a fixed detected energy of neutrinos, Eq.~(\ref{eq:sup}) gives the suppression factor of neutrino flux, which depending on the redshift distribution of sources, can take values even less than $1/2$ for an appropriate value of $\Delta m_j^2$. However, the neutrino telescopes such as IceCube, search for the neutrino flux from cosmological source over a range of energies. Thus, it is also convenient to define the overall suppression factor in the energy range $(E^1_{\nu},E^2_{\nu})$ as

\begin{equation}\label{overall}
\overline{{S}_{\rm eff}(\Delta m_j^2; E^1_\nu,E^2_\nu)}=
\frac{\int_{E^1_\nu}^{E^2_\nu}\sum_k
\left\langle\cos^2\left(\frac{\Delta\Phi_j(z_k,E_\nu^0)}{2}
\right)\right\rangle \frac{dN_\nu
(z_k,E_\nu^0)}{dE^0_\nu}\frac{(1+z_k)}{\left[d_c(z_k)\right]^2}
~{\rm d} E_\nu}{\int_{E^1_\nu}^{E^2_\nu}\sum_k \frac{dN_\nu
(z_k,E_\nu^0)}{dE^0_\nu}\frac{(1+z_k)}{\left[d_c(z_k)\right]^2}~{\rm
d} E_\nu}~.
\end{equation}

The quantity $\overline{S_{\rm eff}}$, for a fixed value of $\Delta m_j^2$, gives a measure of the neutrino flux depletion in the whole energy range $(E^1_{\nu},E^2_{\nu})$.

\end{itemize}

\section{\label{sec:grb}Implications for Neutrino Flux from GRBs}

As discussed in the previous section, depending on the value of $\Delta m_j^2$ and the baseline of the source(s), the flux of neutrinos from cosmological sources at the Earth can be depleted to different degrees. In this section we consider the specific case of GRBs as the sources of high energy neutrinos ($E_\nu\sim$~100 TeV$-$10 PeV). It is shown in \cite{Cecilia} that inside the GRB sources $\int V_{C,N} dt\ll 1$, so the matter effects inside the source are not relevant for neutrino oscillation. On the other hand, inside the Earth we shall have $V_{C,N}\gg \Delta m_j^2/2E_\nu$, so again the matter effect will not be important because of the very tiny effective mixing angle.

\begin{figure}[t]
\begin{center}
\includegraphics[width=0.75\textwidth]{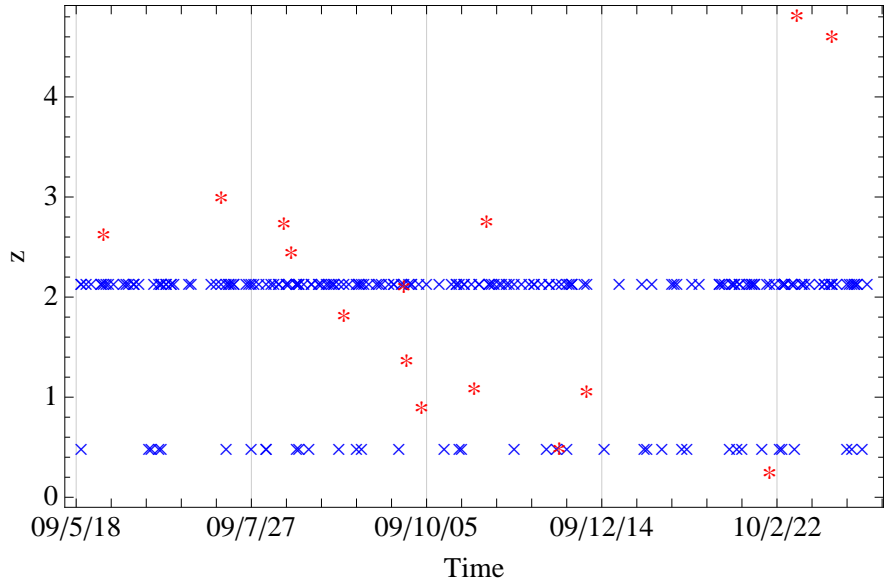}
\caption{\label{fig:grbz}Redshifts of the GRBs observed by their gamma ray emission during the data-taking period of IC59. The red stars show the measured redshifts and blue crosses indicate the assumed values of redshift by the IceCube collaboration. Data taken from~\cite{grbdata}. }
\end{center}
\end{figure}

During the data-taking period of IceCube experiment (with two phases of 40 and 59 strings   out of the completed 86 strings implemented), about 300 GRBs have been observed by their gamma ray
emission. From the X-ray observation of the afterglow it is possible to determine the redshift of the GRBs, which proves the cosmological origin of them~\cite{Kulkarni:2000ux}. However, this
information is not available for all the GRBs. Figure~\ref{fig:grbz} shows the redshift distribution of the observed GRBs for IC59. The red stars show the measured redshifts and blue crosses indicate the assumed values of redshift (by the IceCube collaboration) for GRBs without a measurement of redshift. Although for the majority of GRBs the redshift is not measured, we can generally take $z\sim 1$ so baselines ($L$) are of order of Gpc.

Following the discussion in section~\ref{sec:scenario}, for $L\gg L_{\rm osc}$, the expected flux of neutrinos from GRBs will be reduced by half in the pseudo-Dirac scenario. For the neutrino energy $E_\nu\sim 1$~PeV, this happens for $\Delta m_j^2\gtrsim10^{-16}~{\rm eV}^2$. Thus, taking into account the present upper limit from solar neutrinos~\cite{solar}, mass splitting in the range $10^{-16}~{\rm eV}^2\lesssim \Delta m_j^2\lesssim 10^{-12}~{\rm eV}^2$ will deplete the GRB neutrino
flux by a factor of $1/2$.

Now we discuss the interesting case of $L \sim L_{\rm osc}$. For the GRBs with redshift $z\sim1$ and neutrino energies $\sim$~PeV, this happens for $ 10^{-18}~{\rm eV}^2 \lesssim\Delta m_j^2\lesssim 10^{-16}~{\rm eV}^2$. Following the discussion in section~\ref{sec:scenario}, we first calculate the suppression factor defined in Eq.~(\ref{eq:sup}) for GRB sources. However, as we depicted in Figure~\ref{fig:grbz}, for the majority of GRBs during the data-taking period of IceCube, the redshift is unknown. So, in order to calculate the suppression factor, we take two approaches: $i$) We calculate $S_{\rm eff}$ just for the GRBs with a measured redshift. During the IceCube data-taking time, the redshift has been measured for 48 GRBs. $ii$) We assume that redshift distribution of GRB sources follows that of Star

Formation Rate (SFR). Although in principle at least there is a delay in GRB rate with respect to SFR, to a good approximation the GRB redshift distribution is the same as distribution of core-collapsing massive stars which itself is proportional to SFR distribution ({\it see}~\cite{Porciani:2000ag} for the details of GRB and SFR redshift distribution correspondence). A good fit to the SFR data is given by~\cite{Madau:1999yh}
\begin{equation}\label{eq:dis}
\psi_\ast (z) = C \frac{e^{3.4z}}{45+e^{3.8z}}\frac{\sqrt{\Omega_m(1+z)^3+\Omega_\Lambda}}{(1+z)^{3/2}}~,
\end{equation}
where $\psi_\ast (z)$ is the SFR per unit comoving volume and $C$ is a constant (in the unit of
$M_\odot \, {\rm yr}^{-1} \, {\rm Mpc}^{-3}$). Thus, the number of GRBs with redshift $(z,z+dz)$ is given by $\psi_\ast (z)\cdot dV_c$, where $dV_c$ is the comoving volume element given by
$4\pi D_H d^2_c(z) dz/\sqrt{\Omega_m(1+z)^3+\Omega_\Lambda}$. In order to calculate $S_{\rm eff}$ and $\overline{S_{\rm eff}}$ in Eqs.~(\ref{eq:sup}) and (\ref{overall}), we randomly generated 300 GRBs according to the distribution function $\psi_\ast (z)\cdot dV_c$~. For the energy spectrum of GRB sources, we assumed that the sources at various redshifts have the same luminosity and same spectrum $\propto {(E_\nu^0)}^{-2}$~; {\it i.e.}, we take $dN_\nu(z,E_\nu^0)=N_0 (E_\nu^0)^{-2} dE_\nu^0$ where $N_0$ is independent of $z$. The fluence of neutrinos arriving at Earth from the redshifts between $z$ and $z+dz$ is given by $\psi_* \cdot dV/(4\pi d_L^2)$, where $d_L$ is the famous luminosity distance which is equal to $d_c(1+z)$. It is straightforward to check that $z\simeq 1$ has the largest contribution ({\it i.e.,}   $\psi_*\cdot dV/(4\pi d_L^2)$ has a peak at $z\simeq 1$.) To obtain the average of $\cos^2(\Delta\Phi_j/2)$ term in Eqs.~(\ref{eq:sup}) and (\ref{overall}), we assume a Gaussian distribution of energy with the width $\sigma_E$ to model the energy resolution of the IceCube detector. For the energy resolution, we take two values $\sigma_E=0.1E_\nu$ and $0.5E_\nu$.

\begin{figure}[t]
\begin{center}
\subfloat[]{
 \includegraphics[width=0.5\textwidth]{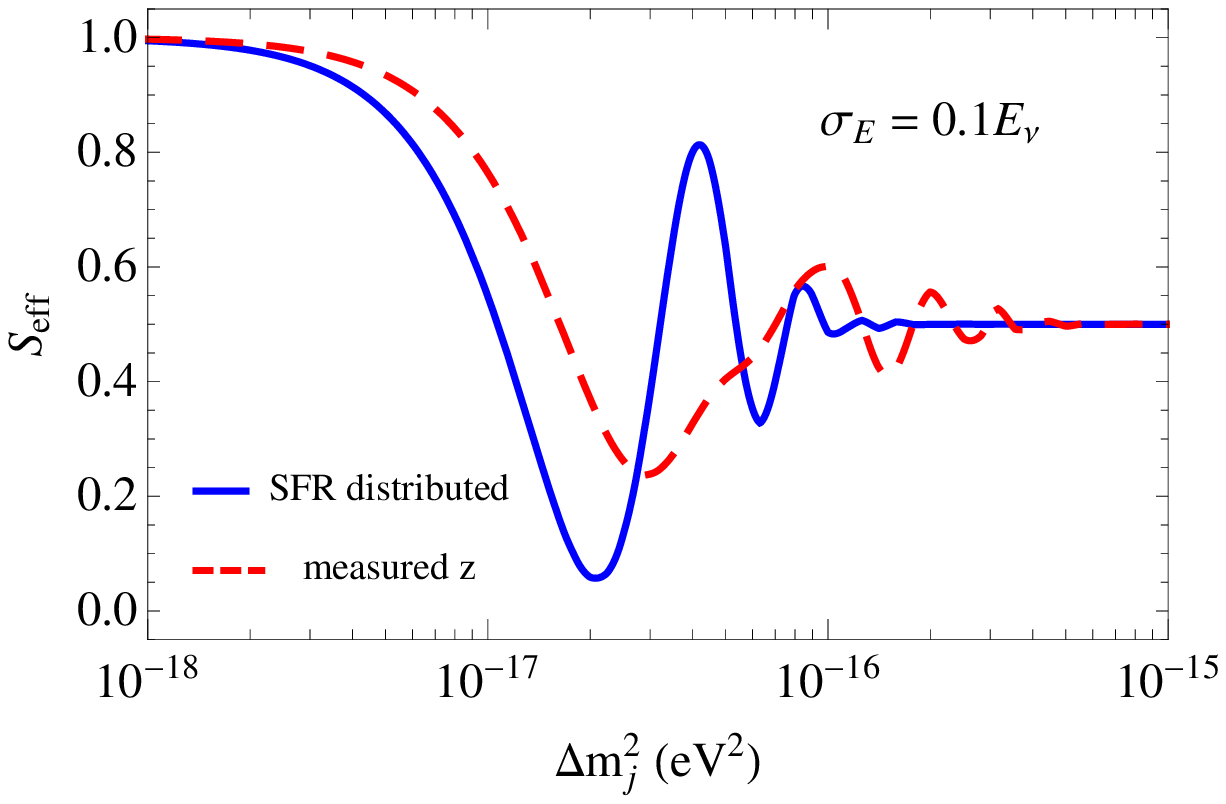}
  \label{fig:seff10}
}
\subfloat[]{
 \includegraphics[width=0.5\textwidth]{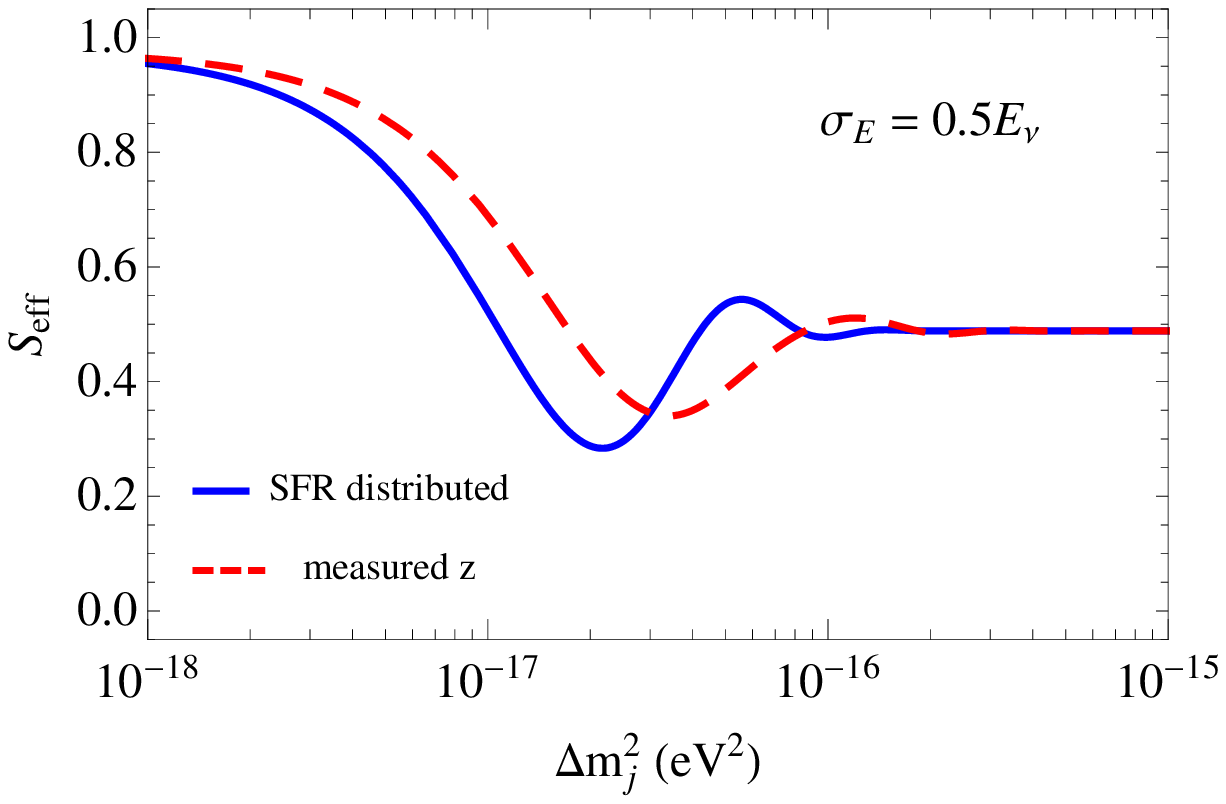}
 \label{fig:seff50}
}
\end{center}
\caption{\label{fig:sup} The suppression factor $S_{\rm eff}$ defined in Eq.~(\ref{eq:sup}) as a function of $\Delta m_j^2$. In the left (right) plot we have taken $\sigma_E=0.1E_\nu$($0.5E_\nu$). In each plot the solid blue curve corresponds to the SFR distributed GRBs and the dashed red curve
corresponds to GRBs with measured redshift. The neutrino energy at Earth is fixed to $E_\nu=1$~PeV.}
\end{figure}

Figure~\ref{fig:sup} shows the suppression factor $S_{\rm eff}$ in Eq.~(\ref{eq:sup}) as a function of $\Delta m_j^2$ for a fixed neutrino energy at Earth $E_\nu=1$~PeV. The plots in Figure~\ref{fig:seff10} and Figure~\ref{fig:seff50} are for $\sigma_E=0.1E_\nu$ and $\sigma_E=0.5E_\nu$, respectively; and in each plot the solid blue curve corresponds to the SFR distributed GRBs and the dashed red curve corresponds for GRBs with measured redshift. For $\Delta m_j^2 \lesssim 10^{-18}\,{\rm eV}^2$, which corresponds to $L\ll L_{\rm osc}$ even for the farthest GRBs, the suppression factor $S_{\rm eff}$ is equal to one, which means that the active neutrinos still do not oscillate to sterile states. For $\Delta m_j^2 \gtrsim 10^{-16}\,{\rm eV}^2$, which corresponds to $L\gg L_{\rm osc}$ even for the nearest GRBs, the suppression factor is equal to 1/2 which have been discussed in Eq.~(\ref{eq:half}). For the intermediate values $10^{-18}\,{\rm eV}^2\lesssim\Delta m_j^2 \lesssim 10^{-16}\,{\rm eV}^2$, the averaging discussed in Eq.~(\ref{eq:half}) do not happen for all the GRBs and thus the suppression factor can have values different from 1/2. Specifically, for $\Delta m_j^2 \simeq 2\times10^{-17}\,{\rm eV}^2$ and $E_\nu =1$~PeV, the flux of neutrinos from SFR distributed GRBs is strongly suppressed; such that for $\sigma_E=0.1E_\nu$, the suppression is almost complete and for $\sigma_E=0.5E_\nu$, the suppression factor is $\simeq 1/3$. The maximum depletion is slightly milder for the GRBs with known $z$ which are more uniformly distributed.

\begin{figure}[t]
\begin{center}
\subfloat[]{
 \includegraphics[width=0.5\textwidth]{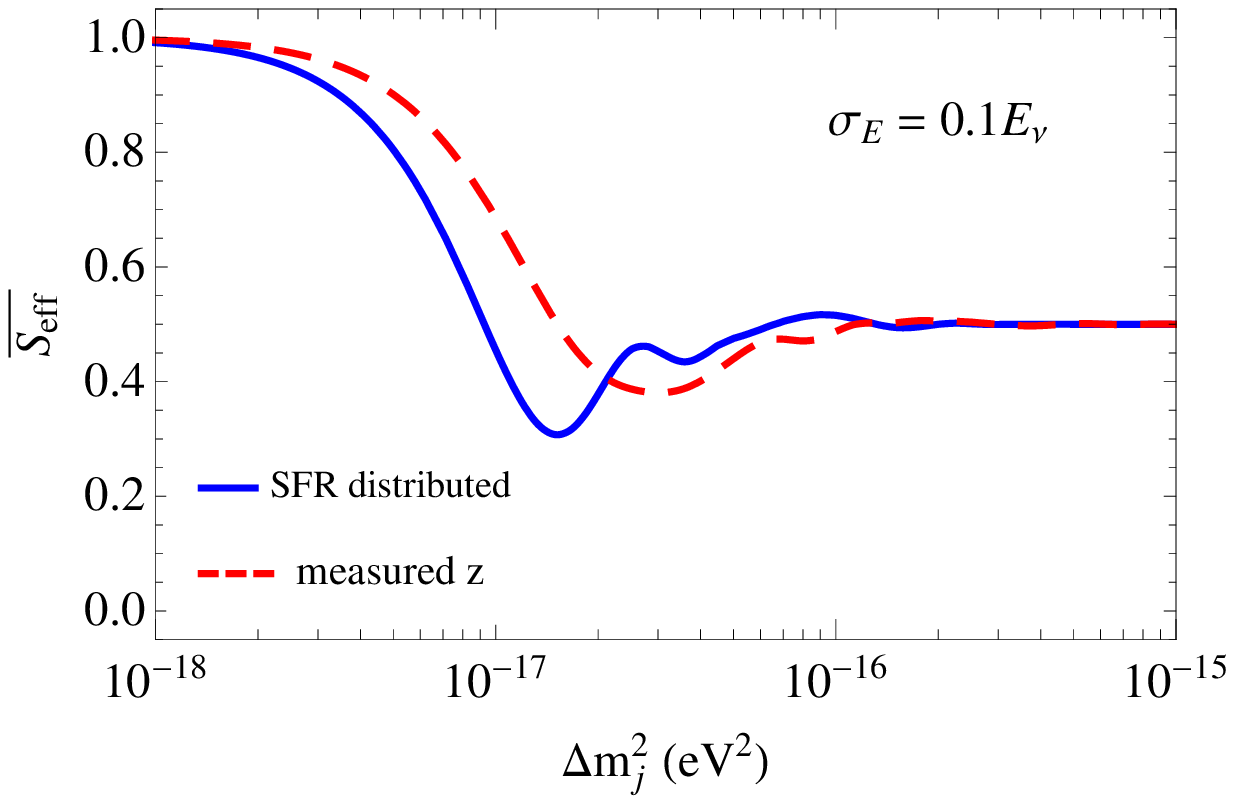}
  \label{fig:seffmean10}
}
\subfloat[]{
 \includegraphics[width=0.5\textwidth]{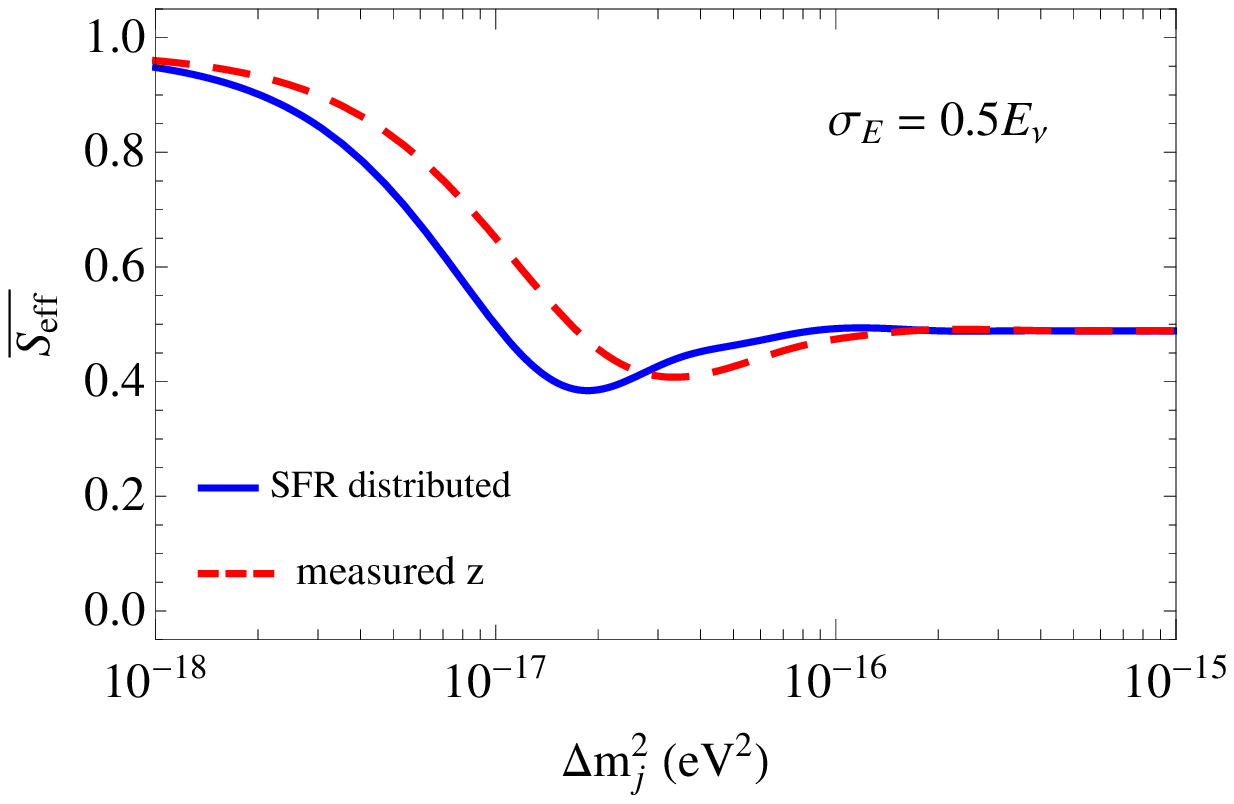}
 \label{fig:seffmean50}
}
\end{center}
\caption{\label{fig:overall} The suppression factor $\overline{S_{\rm eff}}$ defined in Eq.~(\ref{overall}) as a function of $\Delta m_j^2$. In the left (right) plot we have taken $\sigma_E=0.1E_\nu$ ($0.5E_\nu$). In each plot the solid blue curve corresponds to the SFR distributed GRBs and the dashed red curve corresponds to GRBs with measured redshift. The assumed energy range is $(E^1_{\nu},E^2_{\nu})=(0.1,3)$~PeV.}
\end{figure}

The position of dip in Figure~\ref{fig:sup} changes with $E_\nu$. Thus, the strong suppression of the flux at a specific energy does not necessarily imply strong suppression in the whole energy range
of observation in IceCube. We therefore calculate the overall suppression factor in Eq.~(\ref{overall}) to estimate the depletion in the whole energy range. Figure~\ref{fig:overall} shows the $\overline{S_{\rm eff}}$ in Eq.~(\ref{overall}) as a function of $\Delta m_j^2$ for the energy range $(E^1_{\nu},E^2_{\nu})=(0.1,3)$~PeV, corresponding to the energy range of IceCube experiment~\cite{Abbasi}. The color labels are the same as Figure~\ref{fig:sup} and in the left (right) plot we have taken $\sigma_E=0.1E_\nu$($0.5E_\nu$). As can be seen, for $\Delta m_j^2 \simeq 10^{-17}\,{\rm eV}^2$, the overall suppression factor $\overline{S_{\rm eff}}$ can reach $\sim1/3$, with slightly milder suppression for the sample of GRBs with measured redshift.

\begin{figure}[t]
\begin{center}
\subfloat[]{
 \includegraphics[width=0.5\textwidth]{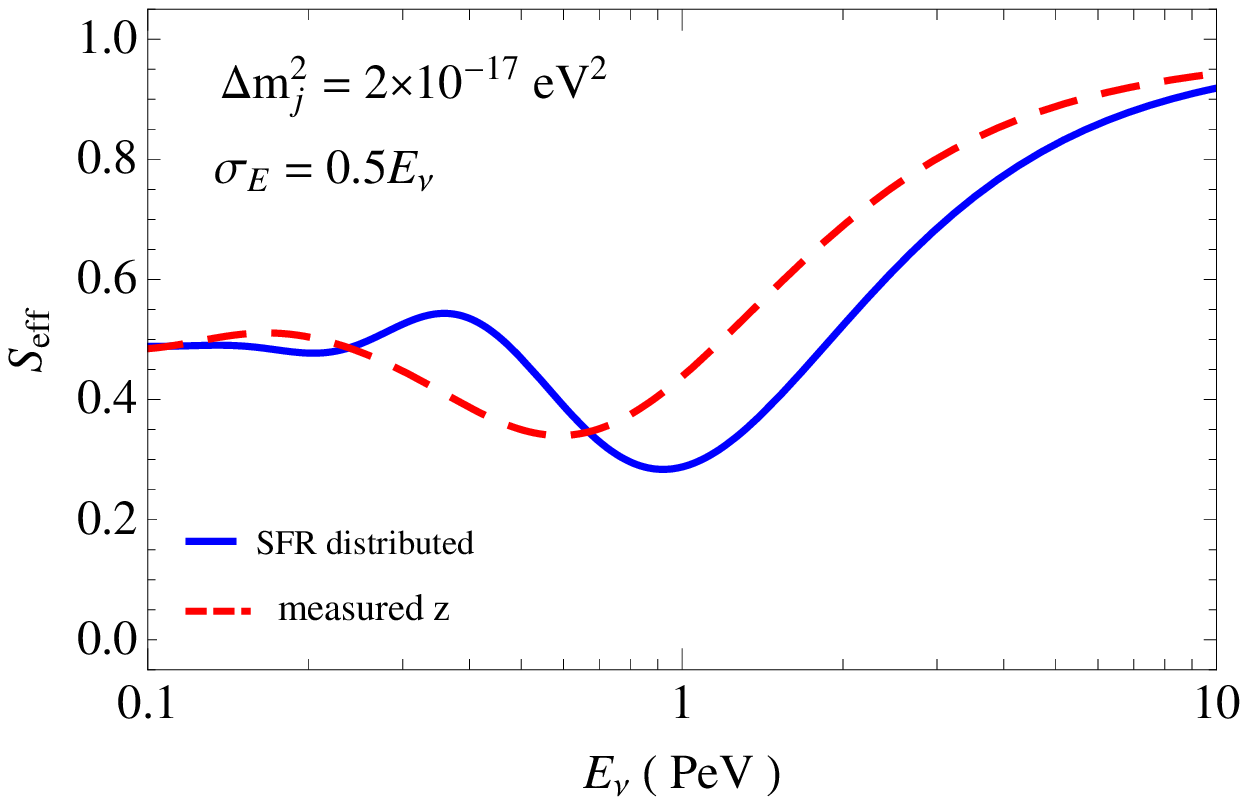}
  \label{fig:seffenergy150}
}
\subfloat[]{
 \includegraphics[width=0.5\textwidth]{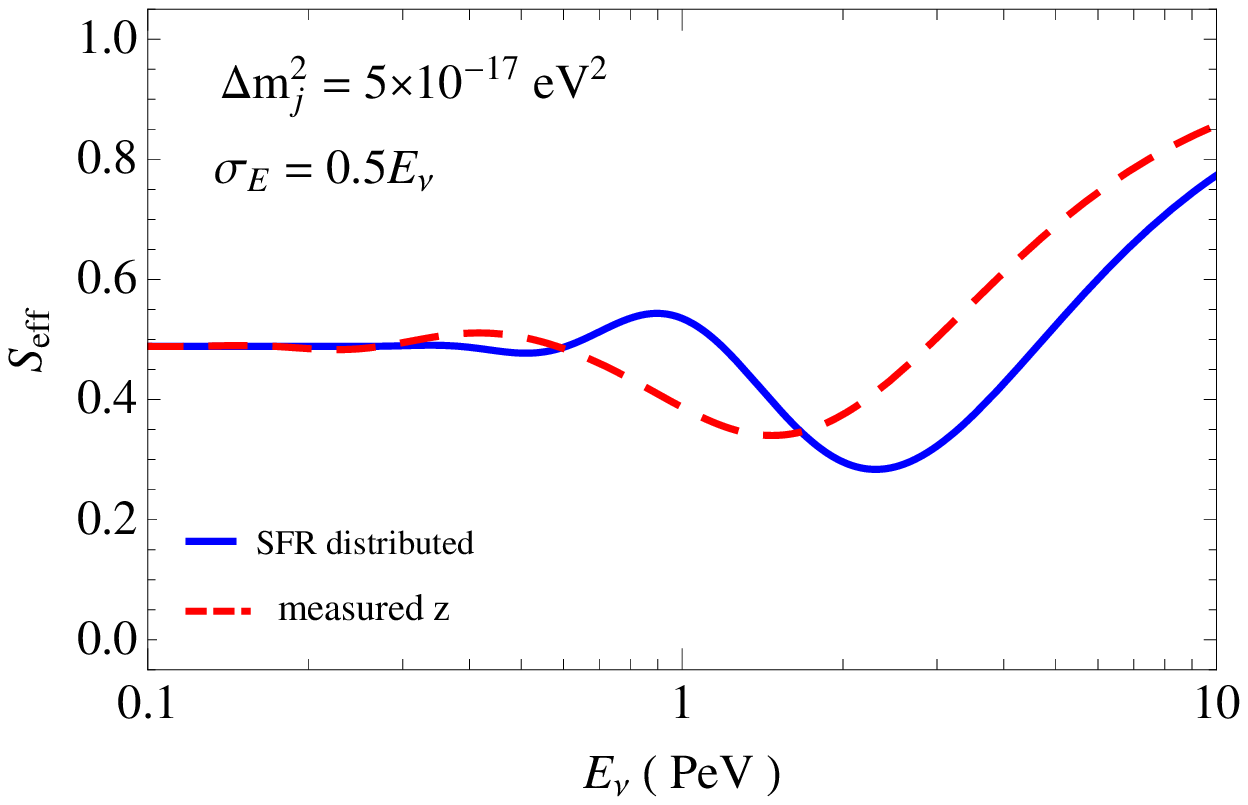}
 \label{fig:seffenergy250}
}
\qquad
\subfloat[]{
 \includegraphics[width=0.5\textwidth]{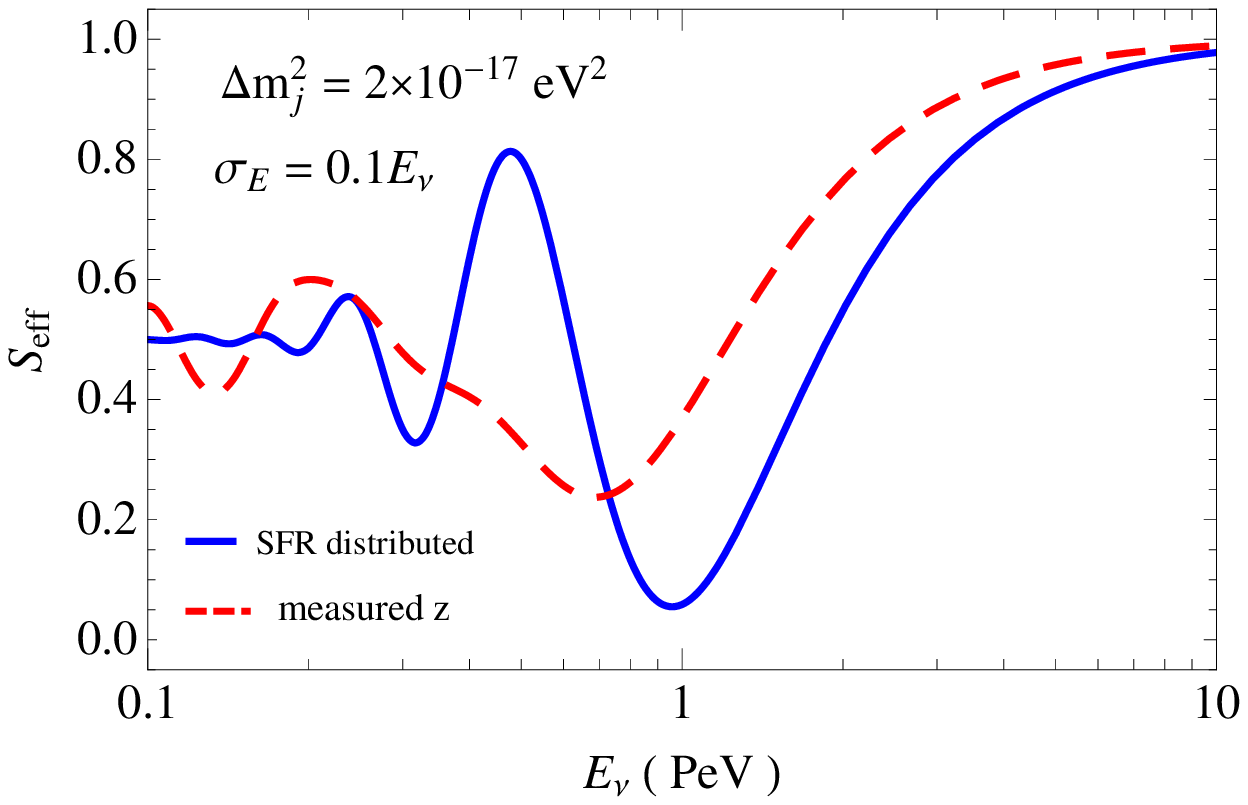}
  \label{fig:seffenergy110}
}
\subfloat[]{
 \includegraphics[width=0.5\textwidth]{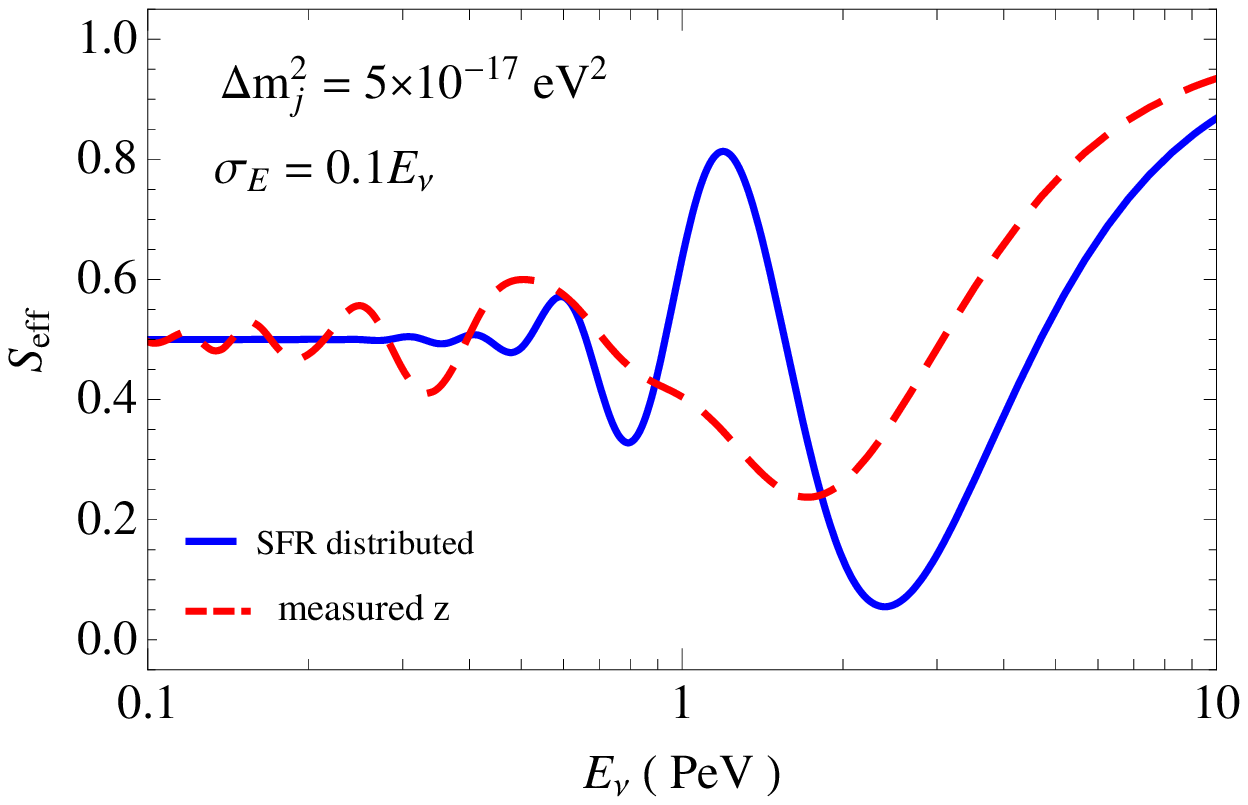}
 \label{fig:seffenergy210}
}
\end{center}
\caption{\label{fig:seffenergy}
The suppression factor $S_{\rm eff}$ defined in Eq.~(\ref{eq:sup}) as a function of neutrino energy for two values of mass splitting $\Delta m_j^2=2\times10^{-17}~{\rm eV}^2$ and $5\times10^{-17}~{\rm eV}^2$, taking $\sigma_E=0.1E_\nu$ and $0.5E_\nu$. In each plot the solid blue curve corresponds to the SFR distributed GRBs and the dashed red curve is for GRBs with measured redshift.}
\end{figure}

In summary, by assuming appropriate value of mass splitting, the active-sterile oscillation in the pseudo-Dirac scenario can lead to strong suppression of the GRB neutrino flux. The suppression can completely deplete the flux for some specific values of neutrino energy. Averaging over the whole energy window of the observation in IceCube it can reach to $\sim1/3$. For the case $\Delta m_j^2\gtrsim10^{-16}~{\rm eV}^2$, the energy spectrum of the GRB neutrino flux is not  distorted and, independent of the neutrino energy, suppression factor is $1/2$. But, for $10^{-18}~{\rm eV}^2\lesssim\Delta m_j^2\lesssim10^{-16}~{\rm eV}^2$, the $S_{\rm eff}$ will have a strong dependence on neutrino energy. To illustrate this point, taking $\sigma_E=0.1E_\nu$ and $\sigma_E=0.5E_\nu$ in Figures~(\ref{fig:seffenergy150},\ref{fig:seffenergy110}) and (\ref{fig:seffenergy250},\ref{fig:seffenergy210}), we show the dependence of $S_{\rm eff}$ on neutrino energy for two values of mass splitting $\Delta m_j^2=2\times10^{-17}~{\rm eV}^2$ and $5\times10^{-17}~{\rm eV}^2$, respectively. In each plot, the solid blue and dashed red curves respectively correspond to the SFR distributed GRBs and to the GRBs with measured redshift. As can be seen, by changing the value of $\Delta m_j^2$ the energy at which strong suppression takes place will change. Comparing the dashed red and solid blue curves in each plot demonstrates that the precise position of the dip in the curves (corresponding to strong suppression in that energy) depends on the redshift distribution of the GRBs. However, the presence of such a peculiar dip in the energy spectrum, independent of the precise position of it, can be interpreted as a hint for the presence of tiny mass splitting $\Delta m_j^2$. As expected, comparing curves for $\sigma_E=0.1 E_\nu$ and $\sigma_E=0.5 E_\nu$ shows that for finer energy resolution the dip is deeper and can be more easily identified. In principle, by reconstructing the energy spectrum of GRB neutrinos in the case of future observation and measuring the GRBs redshift distribution, it can be even possible to determine the values of $\Delta m_j^2$.

\section{\label{sec:sn}Implications for  Supernova type II Neutrinos}

Type II supernovae are the well-known and established sources for neutrinos beyond the solar system. The average energy of the supernova neutrino flux is $\sim 10~$MeV~\cite{Huedepohl:2009wh}. If a supernova explosion takes place within a distance of $\sim 50$~kpc, the neutrino detectors (including Super-Kamiokande~\cite{Ikeda:2007sa}, IceCube~\cite{Abbasi:2011ss} and the proposed liquid Argon experiments~\cite{Autiero:2007zj}) can detect a few thousands of events in a few seconds (pointing towards a point source in case of the possibility to reconstruct the directionality). From Eq.~(\ref{eq:oscL}), for supernova neutrinos the oscillation length will be $L_{\rm osc}\simeq8 \times 10^{-7}~{\rm kpc} \,(10^{-12}~{\rm eV}^2/\Delta m_j^2)$ which is much smaller than the galaxy size or the distance between the only observed supernova up to now (SN1987A) and the Earth. Thus, even for $\Delta m_j^2$ as small as $10^{-18}~{\rm eV}^2$,  we have $L\gg L_{\rm osc}$ and the flux of neutrinos will uniformly be suppressed by a factor of two in the whole energy range of spectrum. As we discussed in section \ref{sec:scenario}, in the averaging limit the suppression factor of $1/2$ will be independent of the possible matter effects inside the supernova.

The expected neutrino flux from a supernova explosion depends on various parameters of the exotic medium inside the explosion region. Based on the supernova explosion models that  are also verified by recent numerical simulations, the neutrino emission from supernova can be described by two phases of accretion and cooling. In the accretion phase the total flux of emitted neutrinos scales with $MT_{\rm acc}^6$, where $M$ and $T_{\rm acc}$ are the accreted mass and the temperature of medium, respectively. In the cooling phase the normalization of flux scales with $R_c^2$ where $R_c$ is the neutrino-sphere radius. Considering the data available from SN1987A, the following values of these parameters and their $1\sigma$ errors can be obtained~\cite{SNmodel}:
$$ M=0.22^{+0.68}_{-0.15}\; M_\odot \quad , \quad  T_{\rm acc}=2.4^{+0.6}_{-0.4}\;{\rm MeV} \quad , \quad R_c=16^{+9}_{-5}\; {\rm km}~. $$
It is easy to see that suppression factor of $1/2$ for the SN1987A neutrino flux can be accommodated within the error of normalization, so it will be a challenging task to investigate an energy independent suppression factor $1/2$ in the total expected number of events.

In principle, the same suppression happens also for the flux of relic supernovae neutrinos (or Diffuse Supernovae Neutrino Flux, DSNF). The DSNF comes from the contribution of all the supernovae explosion at various redshifts, during the history of universe after the era of star formation (see~\cite{Lunardini:2010ab} for a review on DSNF). Up to now, no evidence for such diffuse flux is detected in the experiments. The strongest current upper limit on the flux comes from the Super-Kamiokande analysis of data collected over $\sim$~8 years~\cite{Malek:2002ns}, which is (at 90\% C. L.) $\sim 1.2~{\rm cm}^{-2}~{\rm s}^{-1}$ on the $\bar{\nu}_e$ flux for $E_{\bar{\nu}_e}>19.3$~MeV. Although the DSNF has not yet been observed, the sensitivity of the current experiments is very close to the theoretical predictions and the forthcoming neutrino detectors (such as Hyper-Kamiokande~\cite{Abe:2011ts} and LENA~\cite{Wurm:2007cy}) or new methods of detection (such as doping the Super-Kamiokande with Gd~\cite{Horiuchi:2008jz}) would be able to detect this flux with a good significance.

Theoretically, the DSNF can be calculated by integrating average neutrino flux from a typical supernova weighed with supernova occurrence rate over redshifts $z=0$ to $z\sim4$ which corresponds to the beginning of star formation. The majority of supernova neutrinos from high redshifts, when arrive to detectors, will have energies below the detection threshold, so  the main contribution comes from the supernovae with $z<1$. As discussed above, the normalization of a single supernova flux itself has a large uncertainty. On the other hand, the supernovae rate, which is calculated directly from the supernova observations or indirectly from the SFR, also has a relatively large uncertainty. It is shown in~\cite{Cecilia05} that by taking into account these uncertainties, the theoretical prediction of DSNF can change by a factor of $\sim 4$ at 90\% C.L. Thus, it can be argued that even the observation of DSNF in forthcoming experiment cannot rule out the pseudo-Dirac scenario strictly. However, this scenario can ``rule in'' the models or supernova parameter ranges that predict a flux higher than the present or forthcoming experimental bounds. As discussed before we expect a distortion of the energy spectrum of DSNF for $\Delta m_j^2 \sim 10^{-25}~{\rm eV}^2$ which increases the chance of identifying the active to sterile conversion despite the uncertainties in the total flux. Such small $\Delta m_j^2$ will have no effect on the high energy neutrinos accompanying GRBs.

\section{\label{sec:conc}Conclusions and Discussions}

Within the pseudo-Dirac scenario for neutrino masses, in addition to the two measured mass-squared differences ($\Delta m_{\rm sol}^2$ and $\Delta m_{\rm atm}^2$), there are tiny mass splittings in each pair of active and sterile neutrinos with $\Delta m_{j}^2\lesssim 10^{-12}~{\rm eV}^2$, where the upper limit comes from the analysis of solar neutrinos. In principle, these tiny values of mass-splitting can be probed by neutrinos coming from cosmological sources. The baseline of the neutrinos from cosmological sources with redshift $z\sim1$ reaches $\sim$~Gpc, which means that $\Delta m_j^2\sim10^{-17}~{\rm eV}^2~(E_\nu/{\rm PeV})$ can be probed by the neutrinos with energy $E_\nu$. However, due to the curved nature of space at large scales, the baseline saturates to its maximum value at $z\sim1$ and for neutrinos from sources at higher redshifts, the baseline remains the same. Thus, the $\Delta m_j^2\sim10^{-17}~{\rm eV}^2~(E_\nu/{\rm PeV})$ can be interpreted as the smallest mass-splitting that can be probed by cosmological sources of neutrino with energy $E_\nu$.

We have shown that the active-sterile neutrino oscillation induced by $\Delta m_j^2$ can partially explain the non-observation of the expected neutrino flux from GRB sources in the IceCube detector. The upper limit on the neutrino flux from GRBs derived by the partially deployed IceCube detector is $\sim$~3.7 times stronger than the expected flux, assuming that GRBs are the sources of UHECRs. For $10^{-16}~{\rm eV}^2\lesssim\Delta m_j^2\lesssim10^{-12}~{\rm eV}^2$, the oscillatory terms induced by $\Delta m_j^2$ will completely average out and the neutrino flux from GRBs will be suppressed by a factor of $1/2$~, independent of the neutrino energy. For $10^{-18}~{\rm eV}^2\lesssim\Delta m_j^2\lesssim10^{-16}~{\rm eV}^2$, the averaging does not take place for all the GRBs and the suppression factor can be smaller than $1/2$~. For this case, we have shown that  the average suppression factor over the whole energy range of the IceCube experiment can reach as small as $1/3$, both when we assume that the GRBs are distributed according to the star formation redshift distribution and when we use the redshift distribution of GRBs  for which the redshift is measured (which are less than 20\% of the total number of GRBs studied during the IceCube data-taking period).

With the present knowledge about the explosion mechanism in GRBs, we cannot conclusively rule out/in the pseudo-Dirac scenario for $10^{-16}~{\rm eV}^2\lesssim\Delta m_j^2\lesssim10^{-12}~{\rm eV}^2$. For $10^{-18}~{\rm eV}^2\lesssim\Delta m_j^2\lesssim10^{-16}~{\rm eV}^2$, the suppression factor depends on the neutrino energy. As a result, the energy spectrum of GRB neutrinos will be distorted depending on the value of $\Delta m_j^2$. If the  IceCube detector observes a flux of neutrinos associated to GRBs, by analyzing the shape of the reconstructed spectrum it will be therefore possible to test this scenario.

For the whole range of $10^{-18}~{\rm eV}^2\lesssim\Delta m_j^2\lesssim10^{-12}~{\rm eV}^2$, the oscillatory terms induced by $\Delta m_{j}^2$ completely average out  both for the  point source and diffuse supernova neutrinos. Thus, the flux of supernova neutrinos  will also be suppressed by a factor of $1/2$, independent of the neutrino energy. Due to the large uncertainties in the theoretical prediction of flux normalization, investigating pseudo-Dirac scenario by supernova neutrinos will be challenging. For $\Delta m_j^2\sim 10^{-25}~{\rm eV}^2$, we expect a distortion of the supernova neutrinos energy spectrum whose observation can be a more conclusive hint for the pseudo-Dirac scenario despite the uncertainties in the total flux.

Within the pseudo-Dirac scenario each active neutrino can be written as the sum of two quasi-degenerate mass eigenstates with equal contributions. However, if the scenario is extended such that each active state is the sum of $n$ quasi-degenerate states, in the limit that all the oscillatory terms average out, we obtain a suppression factor of $1/n$.

Throughout the analysis in this paper, for simplicity, we took all the three $\Delta m_j^2$ equal. In fact they can have different values. As long as they are all larger than $10^{-16}~{\rm eV}^2$, the oscillatory terms induced by each splitting will average out and the flux of all the active neutrino flavors will be suppressed by the same factor of $1/2$. If some of $\Delta m_j^2$ are smaller than $10^{-18}~{\rm eV}^2$, the suppression of the neutrino flux from GRB sources will be reduced, as the corresponding state will remain active up to the Earth. If some of the $\Delta m_j^2$ happen to be in the range $10^{-12}~{\rm eV}^2<\Delta m_j^2<10^{-16}~{\rm eV}^2$ and different from each other, the distortion of the energy spectrum will be more complicated than the case that they are all equal.

\begin{acknowledgments}
Authors are grateful to A.~Yu.~Smirnov for useful and stimulating
discussions. Y.~F. acknowledges partial support from the European
Union FP7 ITN INVISIBLES (Marie Curie Actions, PITN- GA-2011-
289442) and  thanks Galileo Galilei Institute for Theoretical
Physics in Florence for its hospitality.
 She is grateful to ICTP for partial financial support and
hospitality. She thanks M.~M.~Sheikh-Jabbari for useful comments. A.~E. would like to thank Pasquale~D.~Serpico, Orlando~L.~G.~Peres and Pedro~C.~de~Holanda for fruitful discussions. A.~E. thanks support from FAPESP.
\end{acknowledgments}

\end{document}